\documentclass[aps,twocolumn,superscriptaddress,nofootinbib]{revtex4}
\usepackage{amsmath}
\usepackage{epsfig}

\newcommand{\bs}[1]{\ensuremath{\boldsymbol{#1}}}
\newcommand{\re}{\ref}
\newcommand{\be}{\begin{equation}}
\newcommand{\ee}{\end{equation}}

\newcommand{\la}{\label}
\newcommand{\ber}{\begin{eqnarray}}
\newcommand{\eer}{\end{eqnarray}}

\begin{document}

\sloppy

\title{ Relation between $(e,e')$ sum rules in $^{6,7}$Li and $^4$He nuclei.\\ Experiment and cluster model}

\newcommand{\Kurchatov}{National Research Center ''Kurchatov Institute'',  123182 Moscow,  Russia}
\newcommand{\mifi}{National Research Nuclear University MEPhI(Moscow Engineering Physics Institute), Moscow, Russia}
\newcommand{\Kharkov}{National Science Center ''Kharkov Institute of Physics and Technology'', 61108, Kharkov, Ukraine}
\author{V.~D.~Efros}\email{v.efros@mererand.com}\affiliation{\Kurchatov}\affiliation{\mifi}
\author{I.~S.~Timchenko}\email{timchenko@kipt.kharkov.ua}\affiliation{\Kharkov} 
\author{A.~Yu.~Buki}\email{abuki@ukr.net}\affiliation{\Kharkov}

\date{\today}

\begin{abstract}
 The sums over $(e,e')$ spectra  of $^6$Li and $^7$Li nuclei which correspond to the longitudinal sum rule
are studied. It is suggested that due to
the  cluster
structure of the lithium isotopes 
these sums 
may  approximately be expressed in terms of such a sum pertaining to the $\alpha$--particle. 
Calculation of these sums is performed
in the framework of
cluster models with antisymmetrization done with respect to all the nucleons.  
At momentum transfers higher than 0.8~fm$^{-1}$ the relations expressing the A=6 or 7 sum in terms of the A=4 sum 
prove to be valid with rather high accuracy. In the region  of momentum transfers around 1~fm$^{-1}$
the longitudinal correlation functions of   $^6$Li and $^7$Li nuclei  are found to be close to that 
of the \mbox{$\alpha$--particle}. Basing on this, the difference between the $q$ values at which the high--$q$
limit of the inelastic sum rule is reached in the $^{6,7}$Li 
cases and the $^4$He case is explained. 
The experimental longitudinal sums in the range between 0.450 and 1.625~fm$^{-1}$ are employed to perform
comparison with the theoretical sum rule
 calculated in the framework of cluster models. Out of the  experimental sums, those in the range between 
0.750 and 1.000 fm$^{-1}$ in the $^6$Li case and between 0.750 and 1.125 fm$^{-1}$
in the $^7$Li case are obtained in the present work.
In the $^6$Li case a complete agreement between experiment and the calculated sum rule
is found while in the $^7$Li case  an agreement only at a qualitative level
is observed.

\end{abstract}

\bigskip

%\pacs{}

%keywords: 
\maketitle

\section{Introduction}

The electronuclear sum rules, see e.g. \cite{orl}, determine two--nucleon correlation functions. In Sect. 2  it is argued
that at momentum transfers around 1~fm$^{-1}$
the longitudinal correlation functions of  $^6$Li and $^7$Li nuclei are close to that of the \mbox{$\alpha$--particle}. 
This originates from the  cluster structure, see e.g \cite{wild}, of
the lithium isotopes. It is argued that
at momentum transfers higher than 0.8~fm$^{-1}$ the
sums over $(e,e')$ spectra  of  $^6$Li and $^7$Li nuclei which correspond to the longitudinal sum rule 
 may be expressed rather accurately  in terms of such a sum pertaining to the $\alpha$--particle.
Calculation of these sums is performed 
in the framework of
cluster models with antisymmetrization done with respect to all the nucleons. 

In Sect. 3  experimental studies at Kharkov Institute of Physics and Technology (KIPT) of  electronuclear reactions with the lithium isotopes are outlined.
Basing on the extracted response functions,  experimental sums corresponding to the longitudinal sum rule
have been obtained in the $^6$Li case at various momentum transfers  lying in the range
 between 0.450~fm$^{-1}$ and 1.450~fm$^{-1}$ \cite{buki77} 
and between 1.125~fm$^{-1}$ and 1.625~fm$^{-1}$ \cite{buki1}.  In the $^7$Li case, such sums have been 
obtained at momentum transfers
in the range between   1.250~fm$^{-1}$ and 1.625~fm$^{-1}$~\cite{buki2}. In the present work, the experimental sums have been 
obtained at several momentum transfers  in the range between 0.750~fm$^{-1}$ and 1.000~fm$^{-1}$ in the $^6$Li case and 
between  0.750~fm$^{-1}$ and  1.125~fm$^{-1}$ in the $^7$Li case. For both nuclei the experimental sums are compared  
with the theoretical sum rule
calculated in 
cluster models.\footnote{Sect. 2 is written by V.D.E. and Sect. 3 is written by I.S.T. and A.Yu.B.}  

The notation below is as follows.
Let $R_L(q,\omega)$ and $R_T(q,\omega)$ be the longitudinal and transverse response functions 
entering the $(e,e')$ cross section. For transitions to continuum one has
\be \frac{d^2\sigma}{d\Omega d\omega}=\sigma_{\rm Mott}\!\left[\frac{Q^4}{q^4}R_L(q,\omega)+
\left(\frac{Q^2}{2q^2}+\tan^2\frac{\theta}{2}\right)R_T(q,\omega)\right].\la{1}\ee
 Here $q$ and $\omega$ are the momentum and energy transfer
from electron to the nucleus, and $Q^2=q^2-\omega^2$. For transitions to discrete levels $R_{L,T}$ are defined in the
same way provided that  
the left--hand side is replaced with $f_{\rm rec}d\sigma/d\Omega$ where $f_{\rm rec}$ is the recoil factor.

We shall consider the longitudinal sums
\be S_{L}(q)=\sum_n\frac{R_{L}(q,\omega_n)}{[{\tilde G}_{E}^p(Q_n)]^2}+ 
\int_{\omega_{\rm thr}}^\infty d\omega\,\frac{R_{L}(q,\omega)}{[{\tilde G}_{E}^p(Q)]^2}
\la{tot}\ee
where the summation goes over discrete levels and includes the elastic contribution. Here and below the notation of the type 
\mbox{${\tilde G}_{E}^{p,n}(Q)=G_{E}^{p,n}(Q)[1+Q^2/(4m^2)]^{-1/2}$} is used, the $G_{E}^{p,n}$ notation stands for 
the electric Sachs proton and neutron 
form factors, $m$ is
the nucleon mass, and $Q_n^2=q^2-\omega_n^2$.

We employ the one--body charge transition operator. One then has the well--known
longitudinal sum rule  
\be S_{L}(q)=Z+\frac{1}{2J+1}\!\sum_{M=-J}^{J}\!\!\langle\Psi_{M}|\sum_{k\ne l}^A{\hat e}_k{\hat e}_le^{i{\bf q}\cdot({\bf r}_k-{\bf r}_l)}|
\Psi_{M}\rangle.\la{sl}\ee
Here $\Psi_{M}$ denotes the ground state with the spin $J$ and its projection $M$, and
\[{\hat e}_k=1/2(1+\tau_{zk})+1/2(1-\tau_{zk})G_{E}^n(q)/G_{E}^p(q)\]
$\tau_{zk}$ being the nucleon isospin projection.
In Eq. (\re{sl}) and below the squared neutron form factor and also the corrections to $G_{E}^n/G_{E}^p$
of the order of $(q/m)^2$
 are neglected. We shall consider only 
moderate $q$ values which justifies \cite{carl} the disregarding of two--body charge operators.  

Comparison with experiment is done for inelastic $(e,e')$ sums so that the
elastic contribution is subtracted from the total sum (\re{sl}). 
In addition, the normalization to unity in the high
$q$ limit is adopted. Thus,
the quantities that are compared with experiment are the inelastic sums 
\ber (S_L)_{\rm inel}(q)=Z^{-1}\{S_L(q)-[ZF_{\rm el}(q)/\tilde{G}_E^p(q)]^2\}\nonumber\\
=1+f_{\rm corr}(q)-Z[F_{\rm el}(q)/\tilde{G}_E^p(q)]^2
\la{inel}\eer
where the correlation contribution $f_{\rm corr}$ corresponds to the second term in Eq. (\re{sl})  and
$F_{\rm el}$ is the charge elastic form factor. 

In experiment of Ref. \cite{buki1} 
it was found that in the $^6$Li nucleus case the $(S_L)_{\rm inel}$ sum approaches its high $q$ limit faster
than for other nuclei. 
%the $q$ behavior of the 
%$(S_L)_{\rm inel}$ sums of 
In this connection let us compare the cases of
$^6$Li and 
$^4$He nuclei. The $^4$He nucleus behaves like many other nuclei in this respect. 
Therefore, an explanation of the 
difference between the two cases
would provide, to some degree,
an explanation of the mentioned feature.
In general, the velocity with which the inelastic sum of Eq.~(\re{inel}) approaches its high $q$ limit   
depends on  the 
velocities of the approaching to zero of the correlation  and the elastic form factor contributions. In the $^4$He case, the second 
of these velocities is lower. Indeed, it is seen from Table 1 in  Ref. \cite{efr1} that   
the inelastic $(S_L)_{\rm inel}$ sum of $^4$He approaches the limit considerably 
slower than the
total  $S_L$ sum does.  
%(This feature may be related to the fact that 
%the average distance between two protons in $^4$He is about $(8/3)^{1/2}$ times larger than
%the size of $^4$He \cite{efr1,efr2}.) 
As a result,
the velocity with which the  $(S_L)_{\rm inel}$ sum approaches its high $q$ limit 
is determined by the elastic form factor contribution {\it i.e.} by the size of
$^4$He. 

And, as argued below, 
 in the $^6$Li case the correlation contribution to the inelastic sum~(\re{inel})  behaves rather similar to that of  $^4$He
as $q$ increases. 
This behavior is due to the contribution of correlations 
between protons belonging to the $^4$He cluster that is a constituent of $^6$Li. 
On the contrary, the absolute value of the
contribution of the elastic form factor to the sum (\re{inel}) decreases considerably faster 
than in the $^4$He case.
This is because of the larger size of $^6$Li. As a result, the $(S_L)_{\rm inel}$ sum  approaches its high--$q$ limit
 faster than in
the $^4$He case. This feature is strengthened by the fact that in the $^6$Li case the correlation contribution 
and the contribution of the
elastic form factor occur to cancel each other to a considerable degree in the corresponding $q$ region.
The said above applies equally to $^7$Li which is a clustered nucleus of a considerable size as well. 
%(In the case of non--clustered nuclei with A>4 
%the  $(S_L)_{\rm inel}$ sum  approaches unity at $q$ values even larger than in the $^4$He case. This is probably caused by a lower effective
%correlation radius for two protons. The limiting case of nuclear matter is exemplified in Ref. \cite{sci}.)
      
\section{sum rules and clustering}

The sum rule (\re{sl}) includes the single--particle term, {\it i.e.}~$Z$, and the correlation term.
Let us consider a limiting case of a clustered nucleus with average distances between clusters much larger than sizes of the clusters.
In such a case, at moderate $q$ values, a
predominant contribution 
to the correlation term comes from regions in the configuration space where clusters do not overlap (see also the reasoning 
below). 
Hence, cluster models  
are sufficient to calculate this term since they
describe properly such regions. Although such models include the operator of antisymmetrization 
with respect to all the nucleons, at the above formulated condition the overlap of  wave function components which differ
in distributions
of nucleons over clusters is not significant. Therefore, the simple cluster model without antisymmetrization
may be employed. 

At these conditions, the following can be suggested.  
Let $r$ be the effective correlation distance for a pair of protons 
belonging to the same cluster, such that the corresponding 
correlation contribution in Eq.~(\re{sl})
becomes negligible when $qr$ is large compared with unity.
 Let $R$ be a
typical distance 
between clusters or between protons belonging to different clusters.
Taking into account that the role of the region of an  overlap between clusters is small one may think that
the correlation distance for protons belonging to different clusters is close to $R$. 
Correlations between such protons 
cease to play a substantial role when  $qR$ becomes large compared with unity.

The correlation distance $r$ anyway cannot exceed the size of a cluster. 
Thus $r\ll R$.
Therefore, there exists a range of $q$ values at which correlations between protons belonging 
to the same cluster are not negligible in Eq.~(\re{sl}) while those between protons belonging to different clusters are negligible.
In this regime, the correlation contribution in Eq. (\re{sl}) can be well represented by the sum of such contributions 
pertaining to the constituent
clusters. 

This relation may
deteriorate  not only at low $q$ values but also at sufficiently
high $q$ values. This is due to antisymmetrization effects disregarded
above. Nevertheless, the following may be stated. 
If one adds $Z$ to this relation then it turns to a relation between the longitudinal sums of a clustered nucleus
and of its constituent clusters. And the latter relation remains approximately valid 
at mentioned higher $q$ values since  
correlation contributions are small at these values.

Suppose that the above picture is applicable in the case of the lithium isotopes. Consider the above mentioned
range of $q$ values for which $qR$ is large compared with unity. 
Then, at such $q$ values, the following relations can be suggested. In the $^6$Li~case,
\be S_L^{^6{\rm Li}}(q)\simeq S_L^{^4{\rm He}}(q)+1.\la{6}\ee
In the $^7$Li~case,
\be S_L^{^7{\rm Li}}(q)\simeq S_L^{^4{\rm He}}(q)+S_L^{^3{\rm H}}(q).\la{7}\ee
(One may think that a similar relation is valid also in the
case of $^9$Be nucleus.)

The cluster model description 
of the lithium isotopes  is used below 
both to estimate the accuracy of these relations and to calculate the $S_L^{^{6,7}{\rm Li}}$ sums. 
Let us start from the case of $^7$Li   that is  a  two--cluster nucleus. First,
let us proceed in the framework of the cluster model without antisymmetrization. 
In this case the ground state wave function of $^7$Li is 
\be \Psi(^7{\rm Li})=\psi_\alpha r_{\rm rel}f(r_{\rm rel})
[\psi_tY_{1}({\bf r}_{\rm rel}/r_{\rm rel})]_{J=3/2,M}\la{str}\ee
where ${\bf r}_{\rm rel}$ is the distance 
between the centers of mass of the two clusters, $r_{\rm rel}f(r_{\rm rel})$ describes the cluster relative motion,
$\psi_\alpha$ is the wave function of the $^4$He cluster, and $\psi_t$ is the wave function of the $^3$H cluster coupled 
with the spherical harmonics to the total momentum. 
The correlation term in Eq. (\re{sl}) consists of three contributions,
the first one corresponds to both nucleons with 
the $k$ and $l$ numbers belonging to the $\alpha$--particle,
the second one to both of them belonging to the tritium, 
and the third one to the case when a nucleon belongs
to the $\alpha$--particle and another nucleon to the tritium. 
The first of these contributions is equal to that of $^4$He. The second one 
is equal to that of $^3$H which is seen if 
one takes into account a simple property of Clebsh--Gordan coefficients. 
Let us denote $\delta$ the third mentioned contribution. To calculate 
it 
we write ${\bf r}_k-{\bf r}_l$
as ${\bar{\bf r}}_k-{\bar{\bf r}}_l+{\bf r}_{rel}$ where  ${\bar{\bf r}}_{k}$ and ${\bar{\bf r}}_{l}$ 
are the nucleon positions with respect to the centers of mass of the clusters.
One obtains that this contribution is as follows,
\be\delta=4[{\tilde G}_{E}^p(q)]^{-2}F_{\rm el}^{^4{\rm He}}(q)F_{\rm el}^{^3{\rm H}}(q) \int_0^\infty\!\!\! dr_{\rm rel}\, r_{\rm rel}^4f^2(r_{\rm rel})
 \frac{\sin qr_{\rm rel}}{qr_{\rm rel}}\la{qq}\ee
where $F_{\rm el}^{^4{\rm He}}$ and $F_{\rm el}^{^3{\rm H}}$ are the elastic scattering form factors. 
 The fact that only the $l=0$ multipole, {\it i.e.} a scalar operator,   contributes to 
the elastic form factor of tritium and a 
simple property of Clebsh--Gordan coefficients are taken into account to get this expression.

We shall use the  $f(r_{rel})$ function from Ref.~\cite{dub}. 
It reproduces accurately all the available scattering and photodisintegration data on various processes involving  $^7$Li. 
The last factor in Eq.~(\re{qq}) becomes very small when $qR$ reaches values large
compared to unity, $R$ being the range of the relative motion function.
To estimate the corresponding $q$ values
let us mention that
the rms value pertaining to the function $rf(r)$ is 3.4~fm. This value  is derived from the values of charge
radii of $^7$Li, $\alpha$--particle, and tritium.  (The $f(r)$ we use leads to the value of 3.7~fm.)

Passing to the net longitudinal sums, one then may write $S_L^{^7{\rm Li}}=S_L^{^4{\rm He}}+S_L^{^3{\rm H}}+\delta$
in the present no--antisymmetrization case.
The contributions  $S_L^{^4{\rm He}}+S_L^{^3{\rm H}}$ and $\delta$ obtained are shown in Table 1 in the second and third column,
respectively. 
 The $S_L^{^4{\rm He}}(q)$ and $S_L^{^3{\rm H}}(q)$
sums are calculated with the help of  the model--independent relations \cite{efr1,efr2} involving
experimental elastic form factors of $^4$He, $^3$He, and $^3$H. These relations have been shown there to be accurate at 
\mbox{$q\le1.5$~fm$^{-1}$}. 
In the $^4$He case
\be S_L^{^4{\rm He}}(q)=2+\left[2+8\frac{G_E^n(q)}{G_E^p(q)}\right]\frac{F_{\rm el}^{^4{\rm He}}(\sqrt{8/3}\,q)}{{\tilde G}_E^p(\sqrt{8/3}\,q)
+{\tilde G}_E^n(\sqrt{8/3}\,q)}\la{4he}\ee
while $S_L^{^3{\rm H}}(q)$ differs from unity merely by a term proportional to $G_E^n(q)/G_E^p(q)$. 
In what follows the 
elastic form factors of $^4$He and  $^3$He from Refs. \cite{ff4,mc}, that of  $^3$H  from Ref.  \cite{ff3} (fit d), 
and the nucleon form factors from Ref.~\cite{nff} are used.

\begin{table}
\caption{The calculated $^7$Li longitudinal sum. Its values without antisymmetrization with respect to nucleons 
belonging to different clusters
 are $S_L^{^4{\rm He}}+S_L^{^3{\rm H}}+\delta$. Its values when antisymmetrization is done are 
presented in the fourth column. In the
fifth column somewhat  improved values (see the text) are listed.  
\label{table1}} 
\begin{center}
\begin{tabular}{|c|c|c|c|c|}
\hline 
$q$, fm$^{-1}$ & $S_L^{^4{\rm He}}+S_L^{^3{\rm H}}$ & $\delta$  &a) $S_L^{^7{\rm Li}}$  
&b) $S_L^{^7{\rm Li}}$  \\
\hline
0.5& 4.60 & 1.96 & 6.50     & 6.58    \\
\hline
0.6& 4.46 & 1.41 &  5.81 & 5.87 \\
\hline
0.7& 4.31 & 0.95 & 5.19 & 5.22 \\
\hline
0.8&4.15 &0.59 &4.66 & 4.68 \\
\hline
0.9&3.99 &0.34 &4.23 &4.24  \\
\hline
1.0&3.84 &0.18 &3.90 & 3.89 \\
\hline
1.1& 3.69&0.085 &3.66 & 3.64 \\
\hline
1.2&3.56 &0.040 &3.47 & 3.45 \\
\hline
1.3&3.44 &0.023 &3.34 & 3.32 \\
\hline
1.4&3.34 & 0.022&3.25 &3.22  \\
\hline
1.5&3.25 &0.025 &3.18 & 3.15 \\
\hline
\end{tabular}
\end{center}
\end{table}

It is seen that the $\delta$ contribution 
is, indeed, relatively small at $q\ge0.8$ fm$^{-1}$. 
At the same time, since the \mbox{$S_L^{^4{\rm He}}(q)+S_L^{^3{\rm H}}(q)$} quantity differs sizably from the $Z=3$ limit 
at corresponding $q$ values, the contribution  to 
the sum  of the correlation term itself is considerable. Thus the relation (\re{7}) is rather accurate.

One also sees that the decrease of $\delta$ at higher $q$ values is not monotonic. This is probably related to the fact that the relative
motion function $f(r)$ has a "shell--model" node at $r\simeq 1.95$ fm that is in the region where the clusters overlap. Due to this, in addition 
to the larger correlation distance $R$ there exists a smaller correlation distance $r_0$ related to the node. This $r_0$ distance is 
such that  $qr_0$ is not so large as compared to unity.
Mainly the $r$ distances smaller than $r_0$ contribute to $\delta$ at higher $q$ values. This contribution is small
since the contribution  to the normalization integral from the distances below the node is 17\% only.

Now, let us estimate the antisymmetrization effects at calculating the correlation function. The $\Psi_M$ 
state in Eq.~(\re{sl})  is
normalized to unity. Let us write  it as
\mbox{$\Psi_M=\langle{\cal A}\psi_M|{\cal A}\psi_M\rangle^{-1/2}{\cal A}\psi_M$ where $\psi_M$} is 
a clustered state corresponding to Eq. (\re{str}) 
and ${\cal A}$ is the antisymmetrization
operator. Below the $\psi_M$ notation stands for the corresponding wave function.
In the calculation below, wave functions of the clusters are chosen to be the products of spatial 
and  spin--isospin functions. Then $\psi_M$ is of the structure  
\mbox{$\psi_M=[\varphi_{L=1}\chi_{STM_T}]_{J=3/2,M}$} where $\varphi_{L=1,M_L}$ are the spatial
factors with given projections of the orbital momentum and $\chi_{SM_STM_T}$ are the
spin--isospin factors. The $\varphi_{L=1,M_L}$ factors consist of  the
spatial components of  wave functions of the clusters and the function of their relative motion.  The
spin--isospin factors consist of spin--isospin functions of the clusters and correspond to
given values of spin and isospin and of their projections.

 The spatial components of wave functions of the clusters we employ are symmetric
with respect to nucleon permutations. They are 
normalized Gaussians of the form \mbox{${\rm const}\cdot\exp[-\lambda\sum_{k=1}^A({\bf r}_k-{\bf R}_{cm})^2]$} where ${\bf R}_{cm}$
is the center of mass variable of a cluster, and $A=3$ or 4. The $\lambda$ parameters are chosen to fit rms matter radii of the clusters.
The  spin--isospin components of these wave functions  are antisymmetric with respect to nucleon permutations.  
The cluster relative motion function   is taken the same as above. It has the form \cite{dub} of a sum of Gaussians.

For brevity let us denote $\varphi_{M_L}$ and $\chi$ the above defined $\varphi_{L=1,M_L}$ and $\chi_{SM_STM_T}$
factors. 
%Let us  write
%${\cal A}\varphi_{m}\chi$ as 
%${\cal A}\varphi_{m}\cdot{\cal A}\chi$.
Since the operator in Eq.~(\re{sl}) does not depend on spin variables, averaging 
over projections of the total spin there turns to averaging over projections of the orbital momentum.
Taking also into account that ${\cal A}^2={\rm const}\cdot{\cal A}$ one may write the correlation term in Eq. (\re{sl}) in the
form
\ber S_L(q)-Z\nonumber=\\ \frac{\sum_{k\ne l}
\sum_{M_L=-1}^1\langle\varphi_{M_L}\chi|{\hat e}_k{\hat e}_le^{i{\bf q}\cdot({\bf r}_k-{\bf r}_l)}{\cal A}|\varphi_{M_L}\chi\rangle} 
{3\langle\varphi_{M_L}\chi|{\cal A}|\varphi_{M_L}\chi\rangle}.\la{aa}\eer  

Suppose that before antisymmetrization
the nucleons with the numbers from one to four   belong to the $\alpha$--particle
and those with the numbers from five to seven belong to the tritium. As a consequence of the fact that
the alpha--particle and the tritium states are already antisymmetric
with respect to nucleon permutations, only the inter--cluster antisymmetrization
is to be done. Furthermore, one can see in this connection that the operator  
\mbox{$1-12(\hat{15})+18(\hat{15})(\hat{26})-4(\hat{15})
(\hat{26})(\hat{37})$}, 
where $(\hat{ij})$ is a particle
transposition, may be put in Eq.~(\re{aa}) in place 
of the total  ${\cal A}$ antisymmetrizer with no change in the result. 

Let us comment on the calculation of the spatial factors in the nominator in Eq. (\re{aa}), for example.
Consider there a single term in the sum over nucleons and one of the 
four
terms with particle permutations.  
The corresponding contribution is an integral over 18 Jacobi variables whose integrand includes a Gaussian times a polynomial.
As noted in Ref. \cite{neu} such type integrals can be done analytically.
In the present calculation, after inserting the additional integration over the  normalized Gaussian wave functions
of the total center of mass, 
the arising total wave function is rewritten in terms of the position vectors of separate particles. This makes simple 
the action of particle permutations. The arising expression is a sum like $\sum_{m\le n}b_{mn}I_{mn}^{kl}$ where
$b_{mn}$ are constants and $I_{mn}^{kl}$ are integrals of the form
\ber I_{mn}^{kl}=
\int d{\bf r}_1\ldots d{\bf r}_N\,
%\left(\prod_{\nu=1}^Nd{\bf r}_\nu\right)
({\bf r}_m\cdot{\bf r}_n)\nonumber\\
\times\exp\left[-\sum_{s,t=1}^Na_{st}\,
({\bf r}_s\cdot{\bf r}_t)+i{\bf q}\cdot({\bf r}_k-{\bf r}_l)\right]
\la{int}\eer
at $N=7$ and $a_{st}=a_{ts}$. 

The integrals (\re{int}) are calculated as follows. As is known
\ber\int dx_1\ldots dx_N \exp\left(-\sum_{s,t=1}^Na_{st}\,x_sx_t+\sum_{j=1}^Ny_jx_j\right)\nonumber\\=\frac{\pi^{N/2}}{\Delta^{1/2}}
\exp\left[\frac{1}{4\Delta}\sum_{s,t=1}^Ny_sy_tA_{st}\right].\la{stint}\eer
Here $\Delta={\rm det}(a_{st})$ and $A_{st}$ is the algebraic adjunct of the $a_{st}$  element
in the $\Delta$ determinant. To calculate the integral whose integrand differs from that in Eq.~(\re{stint}) by the $x_mx_n$ factor
we make a replacement there of the type of
\mbox{$y_i\rightarrow y_i+\lambda\delta_{mi}+\nu\delta_{ni}$}, 
calculate derivatives with respect to $\lambda$ and $\nu$, and then put $\lambda$ and $\nu$ equal to  zero. 
This leads to the following expression for the quantity~(\re{int}),
\ber I_{mn}^{kl}=\frac{\pi^{3N/2}[6A_{mn}\Delta-q^2(A_{mk}-A_{ml})(A_{nk}-A_{nl})]}{4\Delta^{7/2}}\nonumber\\
\times\exp\left[-\frac{q^2}{4\Delta}(A_{kk}+A_{ll}-2A_{kl})\right].\la{for}\eer
($A_{ij}=A_{ji}$.)

At $q=0$ the correlation term in Eq. (\re{sl}) should equal $Z(Z-1)=6$. This is used as a test of the whole calculation. 
(To get this with high accuracy the form factors 
from \cite{nff} at $q=0$ are to be corrected.) And the 
$q$ dependent quantities
(\re{int}) are calculated both using Eq. (\re{for}) and an alternative analytical expression  obtained via differentiating 
the Eq.~(\re{stint}) type equalities
with respect to $a_{st}$. 
 
 The results of the calculation are presented in the fourth column in Table 1. 
 In order to check and somewhat improve them
 let us consider separately the  "direct"\, term {\it i.e.} the contribution  
 that arises when one replaces the $\cal A$ operator with unity in Eq. (\re{aa}). Let us replace this contribution 
 with that calculated as the sum of the second and third columns in the Table. This corresponds to use of the true
 wave functions of the clusters in the direct term in place of those employed above. This is because
  experimental data on form factors of the $^4$He and
 $^3$H nuclei are employed above to calculate the contribution $S_L^{^4{\rm He}}+S_L^{^3{\rm H}}$ and the contribution of Eq. (\re{qq})
 instead of theoretical wave functions of $^4$He and $^3$H.
  The results obtained in this way are presented in the fifth 
 column in the Table. The change in the results is not significant. 
 
 The values of the $S_L^{^7{\rm Li}}$ sum
 without antisymmetrization, {\it i.e.} \mbox{$S_L^{^4{\rm He}}+S_L^{^3{\rm H}}+\delta$}, are rather close to its final values
 obtained. Thus the antisymmetrization effects prove to be small.  And at $q\ge0.9$ fm$^{-1}$ the approximation 
 \mbox{$S_L^{^4{\rm He}}+S_L^{^3{\rm H}}$}
 of Eq. (\re{7}), indeed,  proves to be  close to the net $S_L^{^7{\rm Li}}$ sum.

 If one considers the correlation term itself the antisymmetrization effects are more pronounced.
 They increase as $q$ increases. At $q=0.6$ fm$^{-1}$ the relative deviation of the correlation term
 calculated without full antisymmetrization, {\it  i.e.} 
 \mbox{$S_L^{^4{\rm He}}+S_L^{^3{\rm H}}-3+\delta$}, from that calculated with full antisymmetrization 
  is 0.6\%. At \mbox{$q=0.9$} and 1.1 fm$^{-1}$ it is 6\% and 18\%, respectively, 
 while at $q=1.3$  fm$^{-1}$     
 it reaches already 40\%. However, at higher $q$ values the contribution of the correlation term 
 to the sum is small
 as compared with the $Z=3$ contribution. Even at lower $q$ values the contributions from terms with permutations 
 in Eq. (\re{aa}) prove to be rather
 significant but their effects cancel each other to a large degree. 
 Still,  the term without nucleon permutations is the leading one  which justifies
 the improvement of the results done above.   
 
 In the $q$ region around 1 fm$^{-1}$ 
 the correlation function of $^7$Li proves to be close to that of $^4$He. The latter is well represented  by the second term in Eq. (\re{4he}).
 At lower $q$ values these two correlation functions differ from each other 
 due to correlations between nucleons belonging to different clusters. 
 And at higher $q$ values, where these correlation functions are small, they differ from each other due to 
 antisymmetrization effects.

Let us also comment on the above antisymmetrization procedure. The relative motion function we use was fitted to data in the framework of
the cluster model without antisymmetrization between nucleons belonging to different clusters. And behavior of this function 
in the region of an overlap 
of the clusters was essential at fitting the data.
Our use of this relative motion 
function  in conjunction with the calculation  with full antisymmetrization is an approximation. 
It is still applicable quantitatively as the first approximation 
at those $q$ values at which the effect of inter--cluster antisymmetrization is small.
Such an approximation was used in the literature, e.g. in \cite{eram} in the $^6$Li case.  To our knowledge,
fully antisymmetrized cluster model wave functions of $^{6,7}$Li nuclei fitted directly to representative sets of data are not available
in the literature. (In Refs. \cite{ku,tang}, for example, unsuccessful fits of this type with simple relative motion functions 
are listed.)  Note also that the node of the relative motion function in the region of the cluster overlap reduces antisymmetrization 
effects.  

Next, let us pass to the $S_L$ sum in the $^6$Li case.  The $^6$Li ground state 
wave function is taken to be ${\cal A}\psi$ where $\psi$ is the product of the $\alpha$--particle wave function, 
three--cluster $\alpha+p+n$ relative motion
function, and the spin--isospin function of the outer proton and neutron.  And
${\cal A}$ is the operator of antisymmetrization with respect to  
all six nucleons. 
The $\alpha+p+n$ relative motion
function is taken from Ref. \cite{kuk}. Only its main configuration having the weight more than 95\% is employed.
It is of the form $f(r,\rho)$ where $r$ is the distance between outer nucleons and $\rho$ is the distance between their center of
mass and the $\alpha$--particle.  It is represented in \cite{kuk} as a sum of Gaussians. 
 The correspondent spin--isospin function of outer nucleons has spin equal to one and isospin equal to zero. 

The calculation is done in a way similar to the $^7$Li case above. If antisymmetrization is disregarded then $S_L^{^6{\rm Li}}$ 
can be written as $ S_L^{^4{\rm He}}+1+\delta$ where the $\delta$ contribution comes from correlations between 
outer nucleons and those belonging to the 
$\alpha$--particle. One has 
\ber \delta=4[{\tilde G}_{E}^p(q)]^{-1}F_{\rm el}^{^4{\rm He}}(q)
[1+G_E^n(q)/G_E^p(q)]\nonumber\\\times w^{-1}\int d{\bf r}d{\bs \rho}\,f^2(r,\rho)
\exp[i{\bf q}\cdot({\bf r}/2+{\bs \rho})],\eer
$w$ being the weight of the  $\alpha+p+n$ configuration retained.
The corresponding results are presented in the second and third columns of Table 2. 
The notation is similar to that in Table 1. It is seen that, similar to the  $^7$Li case,  the contribution to $S_L^{^6{\rm Li}}$ 
of correlations between outer nucleons and those belonging to the $\alpha$--cluster is relatively small at $q\ge0.9$ fm$^{-1}$.

\begin{table}
\caption{The calculated $^6$Li longitudinal sum. 
The notation is similar to that in Table 1 except
for the last column where the quantity similar 
to that in the fourth column in Table 1 is presented.
\label{table2}} 
\begin{center}
\begin{tabular}{|c|c|c|c|c|}
\hline 
$q$, fm$^{-1}$ & $S_L^{^4{\rm He}}+1$ & $\delta$  &$S_L^{^6{\rm Li}}$ \\ 
\hline
0.5&4.60 & 1.91 & 6.26 \\
\hline
0.6& 4.45 & 1.38 &  5.58 \\
\hline
0.7& 4.29 & 0.94 & 5.00 \\
\hline
0.8&4.13 &0.60 &4.53 \\
\hline
0.9&3.97 &0.34 &4.17 \\
\hline
1.0&3.82 &0.17 &3.89 \\
\hline
1.1& 3.67&0.055 &3.67 \\
\hline
1.2&3.54 &-0.009 &3.51 \\
\hline
1.3&3.43 &-0.040 &3.39 \\
\hline
1.4&3.33 & -0.051&3.30 \\
\hline
1.5&3.24 &-0.050 &3.23 \\
\hline
\end{tabular}
\end{center}
\end{table}

Furthermore, the calculation of the $S_L^{^6{\rm Li}}$ sum is performed taking 
into account antisymmetrization with respect to all
six nucleons. The same $\alpha+p+n$ relative motion function
is employed as above. The $\alpha$--particle wave function is taken the same as in the 
$^7$Li case above. The results are listed in the fourth column of Table 2. They do not differ much from the 
$S_L^{^4{\rm He}}+1+\delta$ values provided by the no--antisymmetrization approximation. Thus the net  effect of antisymmetrization
proves to be rather small. And at $q>0.9$ fm$^{-1}$ the approximation $S_L^{^4{\rm He}}+1$ of Eq. (\re{6}) 
is, indeed, close to the total $S_L^{^6{\rm Li}}$ sum.

The correlation term entering the $S_L^{^6{\rm Li}}$ sum proves to be rather close to the  correlation term calculated without
antisymmetrization,
especially at lower $q$ values.
At higher $q$ values they are closer to each other than in the $^7$Li case.

The correlation function of $^6$Li proves to be close to that of $^4$He when $q$ ranges between 1.0 and 1.5 fm$^{-1}$. 

At the same time, differently to the $^7$Li case, in the  present case the magnitudes of the direct terms in  the nominator 
and denominator in the relation of Eq.~(\re{aa}) type are comparable with the magnitudes of the terms with permutations. 
Therefore, the improvement
of the results  of the type done in the $^7$Li case is not justified. Also the fact that the net 
effect of antisymmetrization proves to be rather small is curious in view of this feature.
    
\section{Comparison between cluster model  and  experimental sum rules}

The measurements have been performed using the beam provided by the linear electron accelerator LUE--300 at KIPT. 
The primary goal consisted in obtaining the Coulomb energy of the $^6$Li nucleus \cite{buki77} which is 
expressed in terms of the longitudinal sum rule. Later we observed that the inelastic longitudinal sum of $^6$Li 
reaches its limiting value at considerably lower momentum transfers than for other nuclei. In relation to this,
more detailed measurements  have been done in both the $^6$Li and $^7$Li nuclei cases \cite{buki1,buki2}. 

In the last measurements, the beam of electrons with energy ranging from 104 to 259 MeV was employed.
Uncertainty in energy ranged from 0.4 \% to 0.6 \%. The $^{6}$Li and $^{7}$Li targets contained, respectively, 90.5\% and 93.8\% of the
corresponding isotope. Electrons scattered to the angles from $34.2^\circ$ to $160^\circ$ were detected.
Their momenta were analyzed with a double focusing spectrometer \cite{afan}. Electrons were registered in the focal plane
of the spectrometer with an eight channel scintillation--Cherenkov detector \cite{polish1,polish2}. The experimental setup has been described
a number of times in the literature, see {\it e.g.} \cite{{buki02},{buki95}}. A detailed description of the measurements
and data processing is presented in Refs.~\cite{{buki1},{buki2},{buki95},{buku06}}. 

Our task was to extract longitudinal response functions at fixed $q$ values for getting inelastic sums.
Since this required considerable time
the results were published separately for various $q$ values. In the present work, additional 
values of the sums  are obtained, see~Figs.~1 and~2.

\begin{figure}[t]
\begin{center}
\resizebox{0.44\textwidth}{!}{%
  \includegraphics{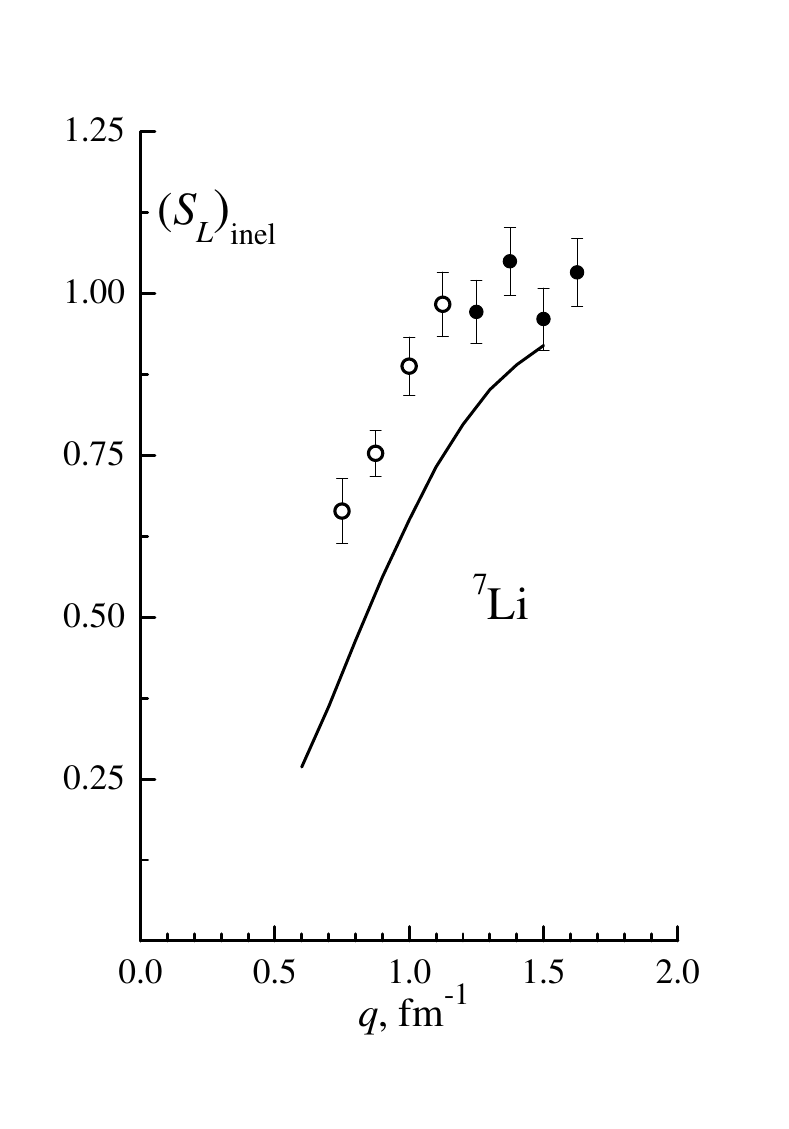}
} \end{center} 
\caption{The inelastic longitudinal sum rule $(S_L)_{\rm{inel}}$ 
of the $^7$Li nucleus. Experimental points from \cite{buki2} (closed circles),
and from the  present work (open circles).
Full curve represents the theoretical values in a cluster model, see the text.}
\label{Fig. 1}      % Give a unique label
\end{figure}

\begin{figure}[t]
\begin{center}
\resizebox{0.44\textwidth}{!}{%
  \includegraphics{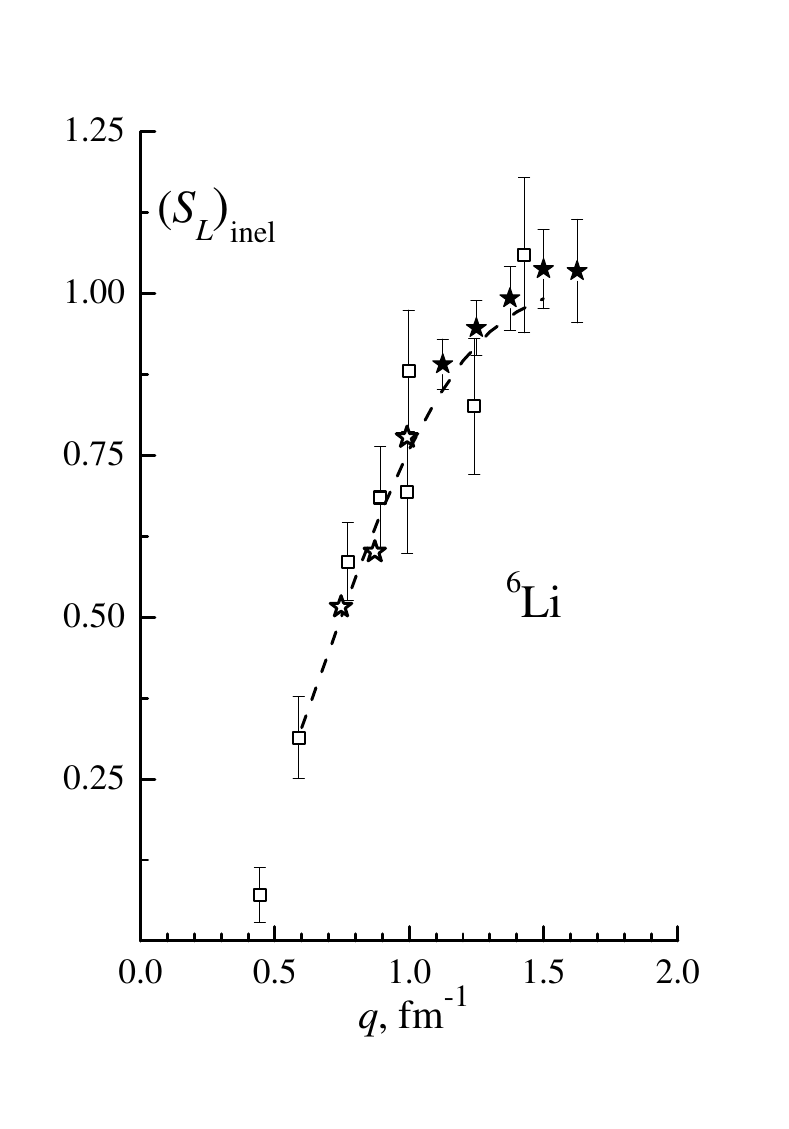}
} \end{center} 
\caption{The inelastic longitudinal sum rule $(S_L)_{\rm{inel}}$ 
of the $^6$Li nucleus. Experimental points from \cite{buki77} (squares), \cite{buki1} (closed stars),
and from  the  present work (open stars without error bars, preliminary). Dashed curve 
represents the theoretical values in a cluster model, see the text.}
\label{Fig. 2}      % Give a unique label
\end{figure} 

To get the sum rule (\re{sl}), the integration over $\omega$ up to infinity is to be performed in Eq. (\re{tot}). 
The $\omega>q$ region of the integrand is kinematically inaccessible in experiment. In fact, the experimental 
response functions $R_L(q,\omega)$ may  reliably be determined 
only at  $\omega$ values considerably smaller than $q$. Therefore, they are to be extrapolated
beyond these values to get the sums. According to Refs. \cite{tor,buki05},
a reasonable extrapolation of the integrand in Eq. (\re{tot}) may be realized with 
a power function $\omega^{-\alpha}$ at $\alpha=2.5$. Up to experimental uncertainties, 
the same $\alpha$~value has been obtained in Refs. \cite{buki1,buki2}. In the present work,
the same power extrapolation at $\alpha=2.5$ was performed. The increase in the 
$(S_L)_{\rm inel}(q)$ sums due to the extrapolation ranged between 8 and 19 per cent.  

Elastic contributions are to be subtracted from 
theoretical sum rules for comparison with experiment, see Eq.~(\ref{inel}). The elastic form factors
of $^{6,7}{\rm{Li}}$ nuclei from Ref.~\cite{1967} are used for this purpose.

In Fig. 1 the inelastic experimental sums $(S_L)_{\rm{inel}}$ of the $^7$Li nucleus are shown along 
with those calculated in the cluster model. The theoretical values are taken from the column 5 of Table 1 with the
elastic contribution subtracted. In addition,  a tiny contribution of the 0.47 MeV excited state is subtracted
since it is not contained in the data. There is only a qualitative agreement between experiment and theory here. 

In Fig. 2 the experimental  $(S_L)_{\rm{inel}}$ sums of the $^6$Li nucleus are compared with those calculated
in the cluster model. The theoretical values are taken from the column 4 of Table 2
with the elastic contribution subtracted.  A complete agreement between experiment and theory is observed.

\section{Conclusion}

Approximate relations expressing the  $(e,e')$ longitudinal sum rules for the $^6$Li and $^7$Li nuclei in terms
of such a sum rule for the $^4$He nucleus are suggested.\footnote{As A. Diaz-Torres noted to us, in general 
such type relations may be of use
to establish whether a nucleus is clustered.} The A=6 and~7  longitudinal sums are calculated in the framework of the cluster models
with antisymmetrization done with respect to all the nucleons. It turns out  that 
at momentum transfers higher than 0.8~fm$^{-1}$ the mentioned relations expressing the A=6 or~7 sum rule in terms of the A=4 sum rule
are rather accurate. Thus the $S_L^{^4{\rm He}}$ sum along with the
sizes of $^6$Li and $^7$Li basically determines the inelastic sums
pertaining to these nuclei.
It is shown
that in the region  of momentum transfers around 1~fm$^{-1}$
the longitudinal correlation functions of the  $^6$Li and $^7$Li nuclei  are close to that of the $^4$He
 nucleus. Basing on this, the difference between the $q$ values at which the high--$q$
limit of the inelastic sum rule is reached in the $^{6,7}$Li 
cases and the $^4$He case is explained.  
 
In the present work, the longitudinal sums are obtained in experiment in the range between 0.750 and 1.125~fm$^{-1}$
in the $^7$Li case and between 0.750 and 1.000~fm$^{-1}$ in the $^6$Li case. (The $^6$Li data are preliminary ones.) 
These experimental sums along with those that were obtained previously at \mbox{$q\le 1.625$~fm$^{-1}$} \cite{buki77,buki1,buki2} 
are 
compared with the sum rule calculated in the framework of cluster models.
In the $^6$Li case a complete agreement is found while in the $^7$Li case  an agreement is only at a qualitative level. 

V.D.E. thanks  S.B. Dubovichenko for useful information.

\centerline{\underline{The running header}: {\it $(e,e')$ sum rules in $^{6,7}$Li}}

\end{document}